\documentstyle[12pt]{article}
\newcommand{\al}{\alpha}
\newcommand{\si}{\sigma}

\newcommand{\fee}{\varphi}

\newcommand{\pa}{\partial}
\newcommand{\be}{\begin{equation}}
\newenvironment{punto}[1]{\begin{itemize}{\item\bf #1\/}:}{\end{itemize}}
\newcommand{\ee}{\end{equation}}
\newcommand{\bea}{\begin{eqnarray}}
\newcommand{\eea}{\end{eqnarray}}
\begin{document}
\begin{titlepage}
\title{Composite Weak Bosons: a Lattice Monte Carlo Analysis}
\author{A. Galli\\[0.2cm]
{\em Paul Scherrer Institute, CH-5232 Villigen PSI, Switzerland}}
\date{\today}
\maketitle
\begin{abstract}
We present a lattice Monte Carlo simulation for the evaluation of the
spectrum of a
confining Yang-Mills theory without Goldstone boson. We show that this
theory is a very good candidate for describing composite weak bosons.
In order to perform the spectrum analysis we have used standard lattice QCD
Monte Carlo
methods. We have also developed an efficient method to evaluate the mass of
the pseudoscalar isosinglet which is present in our theory.
\end{abstract}
\end{titlepage}

\section{Introduction}
The Standard Model \cite{SM} (SM) describes the strong, weak and
electromagnetic
interactions by a gauge theory based on the group
$G_{SM}=SU(3)\times SU(2)\times U(1)$
which is broken by the Higgs mechanism to $SU(3)\times U(1)$. The theory is
essentially determined once the matter fields and their transformation
under the local gauge transformations of $G_{SM}$ are specified. The matter
fields (leptons and quarks) and the Higgs boson are considered to be
elementary. They interact with each other by the exchange of gauge bosons which
are also considered to be elementary. The structure of the SM has been
phenomenologically confirmed to high accuracy.\\
In spite of the beautiful corroboration of the SM by experiments a natural
questions arises: How elementary are the leptons, the quarks, the Higgs bosons
and the gauge bosons?
The idea that the SM itself is an effective theory of another, more
fundamental,
where quarks, leptons and bosons are composites of more fundamental
fields is almost as old as the SM itself. The idea of quark and lepton
compositeness is motivated by the observed connection between quarks and
leptons, by the generation puzzle and by the existence of too many parameters
in the SM. The Higgs compositeness is motivated by the fine tuning problem.
The W and Z compositeness is motivated by their relation to a composite Higgs
and by the observation that all short-range interactions are residual
interactions of a more fundamental long-range interaction.\\
The constituents, the new fundamental fields, are supposed to carry a new
internal quantum number (which we denote as {\em hypercolor}) and the quarks,
leptons and bosons are hypercolorless composite systems of them. The binding of
the
constituents due to hypercolor is viewed as an analogy to the color confinement
mechanism of QCD. However,
since the SM spectrum is different from the hadron spectrum, the hypercolor
interaction has to be described by a strongly coupled Yang-Mills theory
{\em different} from QCD.\\
Several models treat the quarks, leptons and bosons as composite systems.
Today a conspicuous number of theorems
exist which have ruled out most of the existing models and radically restricted
the possibilities to construct realistic composite models.
In principle there are two categories of models:
\begin{punto}{Three-fermion models} In these models the leptons and
quarks have three fermion spin 1/2 constituents. Most of these models are
phenomenologically ruled out \cite{1} by the Weingarten \cite{2}, Nussinov
\cite{3}
and Witten \cite{4}
constraints, by the the 't Hooft's anomaly-matching conditions \cite{5}, by the
Weinberg-Witten theorem \cite{6} or by the Vafa-Witten theorem \cite{7}.
Because most of the constraints dictated by these theorems can be avoided in a
supersymmetric scenario of compositeness,
some authors have proposed supersymmetric versions of these models \cite{8,9}.
\end{punto}
\begin{punto}{Fermion-scalar models} In these models the leptons and
quarks are made out of two constituents, a fermion spin 1/2 and a scalar. The
W, Z and the Higgs bosons can have also two constituents.
Two particular examples are
the "Strongly coupled SM" (SCSM) \cite{10} and the Yang-Mills theory without
Goldstone bosons \cite{11}\footnote{In \cite{11} an unfortunate sign error has
entered the calculations due to a wrong factor $i$ in the fourth and fifth
terms of eq. (3). To interpret the results correctly one has to interchange
isosinglet $\leftrightarrow$ isovector in \cite{11}. I thank H.Schleret for a
discussion of this point.}.\\
The SCSM starts from the same lagrangian of the SM,
however, with an unbroken gauge group $SU(2)\otimes U(1)$. The aim of this
theory is not to describe a model more fundamental than the SM,
but to propose an alternative to the usual Higgs mechanism. The interesting
feature arises that the W and Z are only the lowest lying states of a whole
spectrum of vector-mesons. \\
The Yang-Mills theory without Goldstone bosons is a strongly coupled
non-abelian gauge theory which can have a low lying spectrum equal to the
spectrum of the SM. This is the model we want to study in this work. In
particular we will concentrate on the weak gauge bosons sector comparing the
low lying spectrum of this theory with the SM boson spectrum.
\end{punto}
To be precise this last model considers the photon to remain elementary and
switched off. The weak gauge bosons $W^{\pm}$ and $Z^0$ then form a mass
degenerate triplet.
This model is a usual confining Yang-Mills theory with $SU(2)$ local
hypercolor gauge group, $SU(2)$ global isospin group and generalized Majorana
fermions in the fundamental representation of the local and global symmetry
groups. We note that the generalization of the Majorana fermions with
non-trivial quantum
numbers is possible only if the symmetry groups are real.\\
To be viable  a composite model of the
weak bosons has to reproduce the known weak boson spectrum: the lightest
bound states have to be the W-bosons and heavier bound states have to lie
in an experimentally unexplored energy range.
The only possibility to have a Yang-Mills theory which reproduces the weak
boson spectrum is to choose the degrees of freedom in a way
that they naturally avoid bound states lighter than the vector isotriplet
of the theory which characterizes the W-boson triplet.
This is possible if
the unwanted light bound states which naturally show up as
Goldstone bosons or pseudo Goldstone bosons in many models (like, for example,
a pseudoscalar isomultiplet, which would be the pion analogue of QCD) are
avoided. The choice of Majorana
fermions in this model avoids the $SU_A(2)$ global chiral symmetry of the
Yang-Mills Lagrangian because left- and right-handed degrees of freedom are not
independent. The axial current (which would generate the $SU_A(2)$ chiral
symmetry) does not exist and it is not possible to have a
breaking of $SU_A(2)$ with the related low lying Goldstone bosons.
In fact, the pseudoscalar isotriplet vanishes by the Pauli principle
(it is a symmetric combination of Grassmann variables).\\

Because of the strong coupling character of this theory, we need
non-perturbative methods to make predictions. It is important that the
fermion theory under discussion can be defined by a gauge invariant lattice
regularization. A lattice regularization \`a la Wilson \cite{12}
is possible because the choice of the
isospin group $SU(2)$ allows us to replace the Dirac mass term and the
Dirac-type Wilson term by a hypercolor gauge invariant Majorana type
expression.\\
A strong coupling expansion analysis \cite{13} of the
spectrum of this theory has shown that the spin one isotriplet bound state
(the right quantum number to represent the W-boson of the SM) could be the
lightest state if the pseudoscalar isosinglet acquires a mass by the
chiral anomaly in analogy to the $\eta'$ in QCD. \\
In this work we
calculate the spectrum of the lightest bound states by a quenched
Monte Carlo simulation and we show that the vector
isotriplet bound state of this theory is the lightest one.
We have performed two different type of Monte Carlo simulations.
In the first one we have calculated the masses of bound states without the
contribution of the chiral anomaly. This simulation confirmed the results of
the strong coupling expansion.
In the second one we have developed an efficient method to evaluate the chiral
anomaly
contribution to the mass of the pseudoscalar isosinglet bound state. Its mass
turned out to be heavier than the vector isotriplet mass.\\

Our work is organized as follows: in section 2 we introduce the model in
question and because this paper is not addressed only to lattice specialists
in section 3 we shortly present the technical methods of lattice spectroscopy
that we will use. In
section 4 we explain the method that we have developed to estimate the chiral
anomaly contribution to the pseudoscalar isosinglet mass and in section 5
we analyse the results of the simulations.

\section{Confining Gauge Theories without Goldstone Bosons}
\subsection{The Wilson action}
We consider a gauge theory whose fermion content is represented by a Weyl
spinor
$F_{\al,a}^A(x)$. Here $\al$ denotes the (undotted) spinor index ($\al=1,2$),
$A$
denotes the fundamental representation index of a global SU(2) isospin group
($A=1,2$) and $a$ denotes the fundamental representation index of the local
SU(2) hypercolor gauge group ($a=1,2$). We introduce the generalized
Majorana spinor $\psi$ starting from the Weyl spinors $F$ and its
conjugate $F^\dagger$
\be
\psi(x)=\left(\begin{array}{c}F(x)\\QF^\dagger(x)\end{array}\right)=
\left(\begin{array}{cc}1&0\\0&Q\end{array}\right)\fee(x)
\ee
and its adjoint
\be
\bar\psi(x)=(F^T(x)Q,F^\dagger(x))=\fee^T(x)
\left(\begin{array}{cc}Q&0\\0&1\end{array}\right)
\ee
The matrix $Q$ represents the antisymmetric matrix in spin, hypercolor and
isospin space, which correspond to the Kronecker product of
$i\sigma_2,i\tau_2,iT_2$ (the antisymmetric matrices in spin, hypercolor and
isospin space, respectively). Of course the fields $\psi$ and $\bar{\psi}$ are
not independent fields.
The choice of the global isospin group $SU(2)$ and of the local hypercolor
group
$SU(2)$ allows us to write gauge invariant mass terms for the Majorana
fermion fields
\be
\bar{\psi}\psi=FQF+F^\dagger QF^\dagger
\ee
Note that this choice is unique if one deals with Majorana fermions.
Because of the existence of the mass term we can define the Yang-Mills action
on the lattice in Euclidean space in the form of a Wilson action.
\bea
S&=&\beta\sum_{p\subset\Lambda}TrU(\pa p)-k\sum_{b=\langle
xy\rangle}\bar\psi(x)
\Gamma(b)U(b)\psi(y)-\frac{1}{2}\sum_{x}\bar\psi(x)\psi(x)\equiv\nonumber\\
&\equiv&S_{Gauge}(U)-\bar{\psi} \frac{M(U)}{2} \psi
\eea
For further details we refer to ref. \cite{13}.\\
We will make repeated use of the fact that the fermion fields can be integrated
out from the functional integral. Here, one has to be careful since $\bar \psi$
and $\psi$ are not independent fields and the fermionic functional
integral is represented in terms of $\fee$. As a result we obtain.
\be
Z=\int[d\fee][dU]\exp{\left\{-S(\fee,U)\right\}}
=\int [dU]\sqrt{det\,M(U)}\exp{\left\{-S_{Gauge}(U)\right\}}
\ee
\subsection{CP eigenstates}
In weak interactions CP is a good quantum number but not C and P separately.
Therefore, we perform a classification of the composite operators
according to the CP eigenvalues.
The CP transformation of the Weyl spinor $F$ is defined by
$$
F^{CP}(x)= i\tau_2i\si_2F^{\dagger}(x_P)
$$
(where $x_P=(t,-\vec x)$) which fixes the phase.
In the compact notation we can rewrite the CP
transformation in the following form:
\bea
\psi^{CP}(x)&=&T_2\gamma_0\psi(x_P)\nonumber\\
\bar\psi^{CP}(x)&=&\bar\psi(x_P)\gamma_0T_2
\eea
We can build CP odd and CP even eigenstates\footnote{Lorentz
scalars $S$, vectors $V^\mu$ and tensors $T^{\mu\nu}$ are CP {\bf even} if
$CP\,S=S$, $CP\,V^\mu=V_\mu$ and $CP\,T^{\mu\nu}=T_{\mu\nu}$ and are
CP {\bf odd} if
$CP\,S=-S$, $CP\,V^\mu=-V_\mu$ and $CP\,T^{\mu\nu}=-T_{\mu\nu}$,
respectively.}.
In a confining Yang-Mills theory the physical states are hypercolor singlets.
Because in our model the hypercolor group is $SU(2)$ all physical states are
described by bilinear forms in $\bar\psi$ and $\psi$.
\begin{punto}{Scalar CP eigenstates} The scalar CP {\bf even} combination of
scalars is
\be
S_+(x)=\bar\psi(x)\psi(x)
\ee
The pseudoscalar CP {\bf odd} combination is
\be
S_-(x)=i\bar\psi(x)\gamma_5\psi(x)
\ee
\end{punto}
\begin{punto}{Vector CP eigenstates} The axial vector isosinglet
CP {\bf even} combination is
\be
V^{\mu}(x)=\frac{i}{2}\bar\psi(x)\gamma^\mu\gamma_5\psi(x)
\ee
The isotriplet vector states is
\be
V^{\mu I}(x) =\frac{1}{2}\bar\psi(x)\gamma^\mu T^I\psi(x)
\ee
This state is CP {\bf even}
for I=1,3 and CP {\bf odd} for I=2 and characterizes the W-boson triplet.
\end{punto}
\begin{punto}{Tensor CP eigenstates} The tensors are
\be
\tilde B^{\mu\nu I}=\bar\psi(x)\si^{\mu\nu}T^I\psi(x)
\ee
and their dual
\be
\tilde B^{\mu\nu I}_\ast=\epsilon^{\mu\nu\sigma\rho}\tilde B_{\sigma\rho}^I
\ee
where $\si^{\mu\nu}=\frac{1}{2}[\gamma^\mu,\gamma^\nu]$.
$\tilde B^{\mu\nu I}$ has CP properties opposite to the vector isotriplet $V^I$
ones. Its dual $\tilde B^{\mu\nu I}_\ast$ has CP properties opposite to
$\tilde B^{\mu\nu I}$.
\end{punto}
\subsection{Correlators}
Masses are computed in lattice simulation from the asymptotic behavior in the
Euclidean-time direction of static correlation functions. Static here means
that we have projected the correlation functions at zero momentum transfer.
In our case we need to consider only two
point functions. A typical static bound state propagator can be
written as
\be
C_\Gamma(t)=\sum_{\vec{x}}\langle
\left(\bar\psi(t,\vec{x})\Gamma\psi(t,\vec{x})\right)^\dagger
\bar\psi(0)\Gamma\psi(0)\rangle
\ee
where the $\Gamma$ represent a matrix with hypercolor, isospin and spinor
indices, for example $\Gamma=\gamma^kT^I$ for the vector isotriplet.
Here the contraction of the hypercolor, isospin and spinor indices is implicit.
Contracting creation and annihilation operators into fermion propagators
\be
\langle T\psi(x)\bar\psi(y)\rangle=G(x,y)
\ee
we obtain for the correlation function
\bea
C_{\Gamma}(t)&=&\sum_{\vec{x}}
\langle(\bar{\psi}(x)\Gamma\psi(x))^\dagger
\bar{\psi}(y)\Gamma\psi(y)\rangle =\\
&=&\sum_{\vec{x}}\left\{Tr\left(\Gamma G(x,0)\Gamma G(0,x)-\Gamma
G(x,0)A\Gamma^TAG(0,x)\right)-\right.\nonumber\\
& &-\left.\left(Tr\left[\Gamma G(0,0)\right] Tr\left[\Gamma
G(x,x)\right]-
Tr\left[A\Gamma^TAG(0,0)\right] Tr\left[A\Gamma^TA
G(x,x)\right]\right)\right\}\nonumber
\eea
where $\Gamma^T$ indicates the transposition of $\Gamma$ and A is the matrix
$A=Q\oplus Q$.
Simplifying this expression we obtain the full propagator of the CP
eigenstates.
\be
C_{\Gamma}(t)=
2\sum_{\vec{x}}\left\{Tr\left[\Gamma G(x,0)\Gamma G(0,x)\right]
-Tr\left[\Gamma G(0,0)\right] Tr\left[\Gamma
G(x,x)\right]\right\}
\ee
To prove the formula (15) one has to contract properly all fermion
fields in (13) rewriting them in terms of $\fee$ fields and remembering
that they are not Dirac fields.
The disconnected part of eq. (16) corresponds to the chiral anomaly
contribution to the propagators of the CP eigenstates.
This contribution vanishes for all
isomultiplet bound states. It yields an important contributions
for the pseudoscalar isosinglet bound states $S_-$.\\
The anomaly contribution may be switched off by restricting to the sector of
configurations of topological charge zero.
We will make use of the fact that when the chiral anomaly contribution is
switched off the pseudoscalar isosinglet bound state behaves like a massless
Goldstone bosons and has a propagator of the form
\be
C_{\gamma_5}(t)=
2\sum_{\vec{x}}Tr\left[\gamma_5 G(x,0)\gamma_5 G(0,x)\right]
\ee
On the other hand, when we take into account the chiral anomaly contribution
it is no more a massless Goldstone boson but it acquires a
mass, like the $\eta'$ in QCD. Its propagator is of the form
\be
C_{\gamma_5}^{an}(t)=
2\sum_{\vec{x}}\left\{Tr\left[\gamma_5 G(x,0)\gamma_5 G(0,x)\right]
-\xi N_fTr\left[\gamma_5 G(0,0)\right] Tr\left[\gamma_5
G(x,x)\right]\right\}
\ee
Here $N_f=2$ characterizes the $SU(N_f)$ isospin group.
Notice that we have introduced the parameters $\xi$ in the propagator (18).
This parameter is $\xi=1$ if one computes with dynamical fermions.
Unfortunately, present computational capabilities make necessary the quenched
approximation and the parameter $\xi$ can acquire a value different from
unity \cite{ito}. \\
We will evaluate the mass once without and once with anomaly term. In
the case without the chiral anomaly contribution the mass of the operator
\footnote{To distinguish the two cases we denote by $S_-^{an}$ the pseudoscalar
isosinglet when the chiral anomaly is switched on.}
$S_-$
will be used to identify the chiral limit. The propagator
$C_{\gamma_5}(t)$ can be evaluated using standard Monte Carlo techniques.
For the case where the chiral anomaly contribution is switched on we have
used an improved version of the method of ref. \cite{ito}
to evaluate the propagator $C_{\gamma_5}^{an}(t)$.\\

Because we will evaluate static propagators
we consider the operators being given by the following set
of $\Gamma$'s.\footnote{Because of the relations $\pa_\mu V^\mu=C\times
S_-$
(when the anomaly is switched off)
and $\pa_\mu \tilde B^{\mu\nu A}=C'\times
V^{\nu A}$ (where $C$ and $C'$ are constants) the bound states $V^0$ and
$\tilde B^{0iA}$ are not independent from $S_-$ and, respectively, $V^{iA}$.
Therefore, after summing over all $\vec{x}$ in eq. (13)
we expect that the correlators of $V^0$ and
$\tilde B^{0iA}$ receive contributions from their low-lying states (at large
time separation) and
their masses will be degenerated with the masses of $S_-$ and, respectively,
$V^{iA}$.}\\[0.3cm]
\begin{center}
\begin{tabular}{|c|l|}\hline
$\Gamma$ & flavor \\\hline\hline
$\gamma_5$ & $S_-$\\\cline{1-1}
$\gamma_5\gamma_0$ & \\\hline
$\gamma_5$ & $S_-^{an}$\\\hline
$\gamma_kT^I$ & $V^{kI}$\\\cline{1-1}
$\sigma^{0k}T^A$ & \\\hline
$\gamma_5\gamma_k$ & $V^k$\\\hline
$\sigma^{kj}T^I$ & $\tilde B^{kjI}$\\\hline
1 & $S_+$\\\hline
\end{tabular}\\[0.2cm]
\end{center}
\section{Lattice spectroscopy}
The general
procedure in lattice spectroscopy is to first calculate the
fermion propagator $G(x,y)$ in the presence of an external gauge field and then
to evaluate the hypercolor singlet propagators (16).
The action (4) is quadratic
in the fermion fields, these can be formally integrated out and the fermion
propagator needed to evaluate the correlation functions (16) takes the form:
\bea
G(x,y)&=&\langle T\psi(x)\bar\psi(y)\rangle=
\frac{1}{Z}\int [d\fee][dU]T\psi(x)\bar\psi(0)\exp\{-S(\fee,U)\}=\nonumber\\
&=&\frac{1}{Z}\int [dU]\,M(U)_{x,y}^{-1}\sqrt{det\,M(U)}\exp\{-S_{Gauge}(U)\}
\eea
Just as a simple integral can be evaluated as a limit of a sum, we can evaluate
the path integral (19) by discretizing it on a lattice
in an Euclidian four dimensional space. We put our model on a
$N_s^3\times N_t$ lattice with spacing $a$.
A gauge field configuration is defined as a set of $N\times N$
complex matrices\footnote{$N$ is the number of hypercolors.} defined on each
oriented
lattice bond $b$ with some boundary condition. Fermion fields are defined as
$N\times N_f\times 4$ complex vectors on the lattice points with some
boundary condition.
The lattice spacing $a$ acts
as ultra-violet cut-off and provides a regularization scheme necessary for any
quantum field theory. The Wilson lattice action (4) provides a regularization
which preserves the gauge invariance but breaks the chiral symmetry. The chiral
properties of the theory can be restored in the continuum limit.\\
The calculation of (19) at present is not yet possible because the
computation of $detM(U)$ at each
Monte Carlo step is very CPU time and computer memory consuming.
Like in QCD we assume that the quenched approximation  $det\,M(U)\simeq 1$ is a
reasonable approximation also in our model.
To evaluate the fermion propagator $G(x,y)$ with standard Monte Carlo
techniques
in the quenched approximation
we statistically integrate (19) by replacing the functional integral over
the $U$ variables by the measure $\exp\{-S(U)\}[dU]$ using the standard
algorithms.\\
Given a gauge field configuration $U'$ one obtains the fermion propagator of
this configuration $U'$ by looking at the inverse fermion matrix $M(U')^{-1}$
in eq. (19).
The correlators of the different bound states will be computed by averaging eq.
(16) over different gauge configurations.

\subsection{Matrix inversion technique}
Most of the computational effort computing the correlators goes into
the construction of the fermion propagator $G(x,y)$. We have to find the
inverse of the fermion matrix $M(x,y)\equiv M(U)_{x,y}$, which obeys
\be
\sum_y M(x,y)G(y,z)=\delta_{x,z}
\ee
Since $M$ is a complex matrix
of dimension\footnote{M has lattice indices,
hypercolor indices and spin indices. All isospin index inversion can be
simply done by hand to economize memory and CPU time.}
$V_M= N_s^3\times N_t\times N\times 4$ it is not
possible to store $G(x,y)$ for all $x$ and $y$ since this involves arrays of
the order of $V_M^2$ complex numbers.
Fortunately, we can exploit translation invariance such that one only
needs to know $G(x,y)$ for all
$x$ and for one selected point $y$.\\
Generally, one constructs $G(x,y)$ by solving
\be
\sum_y M(x,y)\tilde G(y,z)=S_z(x)
\ee
where $S_z(x)$ is some external source which can be different from
$\delta_{x,z}$.
The propagator $\tilde G(x,y)$ will be the vector $M^{-1}S$.\\
Because we expect the fermion masses of the constituents
to be much smaller than a
typical bound energy we should invert the fermion matrix for small
masses. However, as the fermion masses tend to the chiral limit,
the matrix inversion of $M$ becomes a hard problem because $M$ tends to
be singular.
Therefore one usually considers a few heavy fermion masses and extrapolates to
the chiral limit. \\
In our calculation the inversion of the fermion matrix is performed
by the Minimal-Residual algorithm or by the Conjugate Gradient algorithm in
the case that the first method does not converge.
To economize computer memory we
organize the inversion algorithms on a checkerboards
even/odd-splitting\footnote{We
defined a lattice point as even (odd) if its coordinate sum $(x+y+z+t)$ is even
(odd). Eq. (21) can be written in a checkerboard basis and
if one knows $\tilde G$ on one checkerboard ($\tilde G_e$, say) one can
reconstruct $\tilde G_o$.}.\\

\subsection{Smearing technique}
The standard method to determine the bound state masses
is to calculate the two-point function
\bea
C_\Gamma(t)&=&\sum_{\vec{x}}\langle O_\Gamma^\dagger(\vec{x},t)O_\Gamma(0)
\rangle\nonumber\\
&=&\sum_n\left\{|\langle n|O_\Gamma(0)|0\rangle|^2e^{-m^\Gamma_nt}+
|\langle 0|O_\Gamma^\dagger (0)|n\rangle|^2e^{-m^\Gamma_n (N_t-t)}\right\}
\eea
where $O_\Gamma(x)\equiv \bar\psi(x)\Gamma\psi(x)$ is an operator with the
quantum
number of the bound state in question, $N_t$ denotes the number of lattice
points in the time direction and $|n\rangle$ is an eigenstate of the transfer
matrix. Here we consider a lattice with periodic boundary conditions.\\
Our goal is to evaluate the mass of the lowest lying state contributing to
(22).
Here the
problem of isolating this term in the sum over all eigenstates occurs.
A reliable value for the mass can be extracted only at large time
t. An efficient method to extract the ground state contribution to $C_\Gamma$
is
to construct an operator $O_\Gamma$ which has a large overlap with the ground
state $|n_0\rangle$. A pointlike
operator $O_\Gamma(x)$ performs very poorly as an operator which generates
bound
states. One therefore constructs a new bound state operator by smearing
\cite{14} the
fermion field $\psi(x)$ with a suitable wave function $F(\vec{x},\vec{x}')$.
One has to choose the
wave function which has a physical extent over several lattice spacing. $F$ has
to be smooth and to vanish for large separation $|\vec{x}-\vec{x}'|$.\\
We have chosen $F$
dependent in a gauge invariant way on the link gauge variables $U$. The smeared
fermion field takes the form:
\be
\psi_{s}(\vec{x},t)=\sum_{\vec{x}'}F(\vec{x},\vec{x}',U,t)\psi(\vec{x}',t)
\ee
where the subscription "s" indicates smeared fields.
In this work the wave function is chosen to be of the Gaussian type
(characterized by two parameters $\al$ and $n$)
\be
F(\vec{x},\vec{x}',U,t)=\left(1+\alpha
H(\vec{x},\vec{x}',U,t)\right)^n
\ee
with the hopping matrix
\be
H(\vec{x},\vec{x}',U,t)=\left\{U(\langle xx'\rangle)+
U^\dagger(\langle x'x\rangle)\right\}|_{x=(t,\vec{x}),x'=(t,\vec{x}')}
\ee
This wave function is defined in local time which means that it involves gauge
fields only in one time slice and is gauge covariant.\\
The smeared fermion propagator $\tilde G_s(x,x')$ is related to the local
fermion
propagator $\tilde G(x,x')$ by
\be
\tilde G_s(x,x')=\sum_{\vec{y}}F^\dagger(\vec{x},\vec{y},U,t)\sum_{\vec{z}}
\tilde G(y,z)F(\vec{z},\vec x',U,t)
\ee
where $x=(t,\vec x)$ and $x'=(t',\vec x')$.
We mention that the construction of the smeared fermion propagator does not
take much more computer time than the calculation of the local propagator.\\
The ground state dominance of the correlation functions $C_\Gamma$ is signalled
by the occurrence of a plateau in the {\em local mass}
\be
\mu_\Gamma(t)=\log\left(\frac{C_\Gamma(t)}{C_\Gamma(t-1)}\right)
\ee
In this case the local mass
is time independent for a range of $t$. It was impossible to attain a plateau
without smearing the fermion fields. In our calculation we have tuned the
parameters $\alpha$ and $n$ of the wave function (24)
to obtain early plateaux in the local masses of the smeared operators.
Defining
\be
r^2=\left\langle\frac{\sum_{\vec x}\vec x^2Tr[F(\vec x,\vec 0)F^\dagger
(\vec x,\vec 0)]}{\sum_{\vec x}Tr[F(\vec x,\vec 0)F^\dagger
(\vec x,\vec 0)]}\right\rangle
\ee
the optimisation of the wave function coincides in our calculation
with the radius $r$ approximately of $3\times a$,
where $a$ is the lattice spacing.
This is a resonable size for a bound state of a typical size less or equal
the spacing $a$.

\subsection{Setting the scale and the hopping parameter}
{}From lattice calculation we can evaluate only dimensionless quantities.
To connect
these quantities to physics we have to set a scale. On the lattice we
choose the scale to be the lattice spacing $a$ and we express all quantities in
unit of $a$. For small enough $a$ the lattice results up to a renormalization
of the parameters and the fields yield the continuum theory.\\
In our model the experimental value of the W-boson mass
will be used to setting the $a$ scale.
For setting the space lattice $a$ we
identify the experimental value $M_W=80.22(26)$ GeV with the value of the mass
of the $V^{kA}$ resonance in the dimensionless quantity
$m_{\gamma^kT^A}\times a$ determined from the simulation.\\
The lattice value of $m_{\gamma^kT^A}\times a$ has to be determined for
heavy fermion masses to avoid the singularity of the fermion matrix.
However, we expect that the constituent fermion masses are much smaller than
a typical binding energy.
Therefore we may proceed in analogy to QCD and extrapolate the
quantities to the chiral limit.\\
In QCD the critical value $k_c$ of the hopping parameter follows from the
determination of the pion mass
\be
a^2\times m^2_{\pi}=A\left(\frac{1}{k}-\frac{1}{k_c}\right)
\ee
suggested by the lowest order chiral perturbation theory \cite{15}.
The pion is the
Goldstone boson  predicted by the spontaneously broken chiral symmetry and
should be a massless
particle in the chiral limit. The non-vanishing experimentally measured
mass of the pion is the result of the breaking of flavor SU(2) symmetry. \\
In our model there is {\em no}
Goldstone boson which can play the role of the pion in QCD.
However, when the chiral anomaly is switched off
the pseudo scalar bound state behaves like a Goldstone boson.
We emphasise that swiching on the chiral anomaly
by choosing topological {\em non} trivial gauge configurations
we expect that the pseudoscalar bound state $S_-^{an}$ acquires a mass
by the chiral anomaly in analogy to the $\eta'$ in QCD. \\
The critical value $k_c$ of the hopping parameter can be evaluated by assuming
that the pseudo scalar bound state $S_-$ behaves like a
Goldstone boson and therefore it should be massless.
In analogy to QCD one makes the following ansatz
\be
a^2\times m^2_{\gamma_5}=A\left(\frac{1}{k}-\frac{1}{k_c}\right)
\ee
In our fits the
linear dependence of $m^2_{\gamma_5}$ on the inverse
hopping parameter $1/k$ is very well satisfied.
All other bound state masses can be evaluated by linearly extrapolating their
lattice predictions $m_\Gamma\times a$ to the critical hopping parameter $k_c$
as suggested by our strong coupling expansion \cite{13}.
In particular, for the vector isotriplet $V^{kA}$ mass, the
dimensionless quantity $m_{\gamma^kT^A}\times a$ which is used to set the
scale can be extrapolated into the chiral regime by the
linear ansatz
\be
a\times m_{\gamma^kT^A}=B+C\left(\frac{1}{k}-\frac{1}{k_c}\right)
\ee
in analogy to the linear ansatz for the $\rho$ mass in QCD.

\section{The pseudoscalar isosinglet mass.}
We have shown in the ref. \cite{13} that the vector isotriplet bound state can
be the
lightest bound state provided that the pseudoscalar isosinglet acquires a
mass from  the chiral anomaly.
In order that our model may be considered a viable theory of electroweak
compositeness it has to turn out that
this bound state is heavier than the vector isotriplet one. \\
A non-perturbative estimate of the contribution of the
chiral anomaly to the pseudoscalar isosinglet mass is one of the
most challenging problems of the lattice Monte Carlo simulations.
In the past there were a few attempts to make a quantitative
Monte Carlo analysis of that problem in QCD \cite{ito,fuk,Q}.
In this section we will discuss the calculation of the mass of the pseudoscalar
isosinglet bound state by applying and improving, using the smearing
technique, some methods developed in \cite{ito,fuk}.\\
In QCD, in the limit of zero bare masses of the u and d quarks,
the pseudoscalar mesons ($\pi$ and $\eta'$) are
Goldstone bosons due to the spontaneous symmetry breaking of the
SU(2) axial symmetry.
Because the quarks possess small but nonzero bare masses,
the $\pi$ has a small mass and only approximates a Goldstone boson.
However the $\eta'$ meson is too heavy for approximating a Goldstone boson.
This is the $U(1)$ problem \cite{thoof}.
t'Hooft removed the puzzle by showing that the existence of topologically
non-trivial gauge configurations such as instantons resolves the
$U(1)$ problem.\\
In any confining Yang-Mills theory with fermion carrying non-trivial
isospin quantum number the same problem arises. The pseudoscalar isosinglet
bound states (the analogous to the $\eta'$ in QCD) acquire heavy masses due to
instanton effects.Their masses can be evaluated using the
method of lattice gauge theories. There are, however, a number of technical
difficulties which restrict the feasibility of the study.\\
The propagator of the pseudoscalar isosinglet $S_-^{an}$ bound state is given
by
equation (18).
{}From the numerical evaluation of this propagator we can determine the mass
of the $S_-^{an}$ bound state
using the lattice spectroscopy methods presented in section 3.
The major difficulty
comes from the large amount of computer time needed to evaluate the
disconnected
contribution to $C_{\gamma_5}^{an}$.
For this purpose we need the fermion propagator $G(x,x)$ for all
points $x$ , which requires $N_s^3\times N_t$ inversions of the
fermion matrix using the method of the point source on a lattice of size
$N_s^3\times N_t$. This is a hopeless task because it needs to much computer
time.\\
The problem can be solved requiring $N_t$ fermion matrix inversions by
calculating the ratio
\be
R(t)=\frac{C'^{an}_{\gamma_5}(t)}{C'_{\gamma_5}(t)}
\ee
for $N_t$ walls .
Here, $C'_{\gamma_5}$ and $C'^{an}_{\gamma_5}$ denote the propagators computed
at a suitable chosen value of $\vec x$ in eq. (17) and (18). This means that
one does not
perform the sum over $\vec x$ in (17) and (18).
The long distance behaviour of the propagators (17-18) for a
fixed $\vec x$ is then \cite{frol}
\bea
C'^{an}_{\gamma_5}(t)&=&
A\left(\frac{e^{-m_{\gamma_5}^{an}t}}{t^D}+
\frac{e^{-m_{\gamma_5}^{an}(N_t-t)}}{(N_t-t)^D}\right)\nonumber\\
C'_{\gamma_5}(t)&=&
B\left(\frac{e^{-m_{\gamma_5}t}}{t^{D'}}+
\frac{e^{-m_{\gamma_5}(N_t-t)}}{(N_t-t)^{D'}}\right)
\eea
where $A,B$ and $D,D'$ are constants, $m_{\gamma_5}^{an}$ and  $m_{\gamma_5}$
denote
the masses of $S_-^{an}$ and $S_-$, respectively.
The
smearing of the source over the space-like lattice for each time wall
has the effect to partially sum over the space-like point over each
time wall and thus reduces the $t^D$ and $t^{D'}$ dependence of the
propagators (33). To a good approximation we expect that with smearing the
exponents $D$ and $D'$ are $D\simeq D'\ll 1$.
The ratio $R(t)$ eliminates the $t^D$ and $t^{D'}$ dependence of the
propagators
$C'_{\gamma_5}$ and $C'^{an}_{\gamma_5}$ and can be expressed
for sufficiently large time t by
\be
R(t)\simeq E\left(e^{-\Delta m_{\gamma_5}t}+e^{-\Delta
m_{\gamma_5}(N_t-t)}\right)
\ee
where $E=A/B$ is a constant
and $\Delta m_{\gamma_5}=m_{\gamma_5}^{an}-m_{\gamma_5}$
represents the mass difference between the
$S_-^{an}$ and the $S_-$ bound states for any $k\leq k_c$. Because, in the
quenched approximation, this mass
difference is a pure hypergluonic effect, it
does not depend on the mass of the Majorana fermion fields and
also on the hopping parameter $k$.
At the chiral limit
the mass of $S_-$ is equal to zero and $\Delta m_{\gamma_5}$ correspond to the
mass of
$S_-^{an}$. Checking that the ratio
$R(t)$ is independent of the hopping parameter $k$ one can
extrapolate $\Delta m_{\gamma_5}$ to the chiral limit by a constant function in
$k$ obtaining then $m_{\gamma_5}^{an}$.\\
The contribution of the disconnected term to the propagator of $S_-^{an}$ is
present only if topological nontrivial gauge configurations are
used. $C'^{an}_{\gamma_5}= C'_{\gamma_5}$ (statistically)
for topological trivial configurations. Topological nontrivial configurations
can be obtained thermalizing an instanton on the lattice using the heat bath
updating. The cooling algorithm is also used to determine the topological
property of the configurations. Cooling \cite{cool} works as described next:
we use the
standard plaquette action which will be locally minimized during the iterations
sweeps of the heat bath algorithm.
The effect is to locally smoothen the lattice gauge configuration,
removing the ultraviolet fluctuations which are responsible for destroying the
topological charge.
The starting instanton
which will be thermalized can be defined on the lattice by discretizing the
instanton solution proposed by t'Hooft \cite{thoof}. To control the topological
properties of the gauge configurations we have used Peskin's lattice
definition of the topological charge \cite{pesk} in the symmetrized version
\cite{pesk2}
\be
Q=-\sum_{x\in\Lambda}\sum_{(\mu,\nu,\rho,\sigma)=\pm 1}^{\pm 4}
\frac{\tilde\epsilon_{\mu\nu\rho\sigma}}{2^432\pi^2}Tr\left[U(x)_{\mu\nu}
U(x)_{\rho\sigma}\right]
\ee
which has the right continuum limit.
Here $\tilde\epsilon_{\mu\nu\rho\sigma}$ is the generalized total
antisymmetric tensor\footnote{For example:
$1=\tilde\epsilon_{1234}=-\tilde\epsilon_{-1234}$.}
in any direction (positive and negative) of the Euclidean space
and $U(x)_{\mu\nu}$ is the product of four links $U(b)$ around a plaquette
lying in the ${\mu\nu}$ plane and with starting point $x$. There are other
definitions of the topological charge on the lattice \cite{Q,Lus}. We have
chosen
Peskin's one because it is the simplest to be programmed on the computer. We
emphasize that a less naive (and more complicate)
definition of the topological charge would be needed
if one would like to know the topological charge density $Q(x)$
or the topological susceptibility like in \cite{Q}.
However, in our work we only need a naive
control over the global topological charge $Q=\sum_{x\in \Lambda}Q(x)$ and the
simple definition of Peskin is sufficient for our purpose.\\
In order to perform a Monte Carlo integration with topological non-trivial
configurations
one has to generate a set of configurations having integer topological charges.
Unfortunately, because the lattice definition (35) of the topological charge
is a discretization of the continuum definition which allows non integer values
of $Q$, we can't generate configurations with
an exact integer topological charge, but only with a $Q$ lying
near an integer value. We accepted only
configurations with $Q$ lying in an interval $\pm 0.1$ around an integer
number.\\
An other difficult is to generate configurations with $|Q|> 1$ because the
small volume of the lattice does not allow a proper thermalization of such
instantons: they loose their topological charge during the iterations
and and in configurations with $|Q|=1$ or $Q=0$.

\section{Results}
\subsection{Computational details}
The simulation was performed on the Cray YMP at the ETH in Z\"urich and on the
NEC SX-3 at the CSCS in Manno.
The simulation was done on different lattices in order to understand the finite
volume and $a$ effects (see Table 1).
The $SU(2)$ configurations were generated by the combination of
heat bath and over-relaxed updating (1 heat bath for 6 over-relaxed sweep).
The cooling updating was also used for topological non-trivial configurations
to determine the topological charge.\\
The plateaux of the local masses (27) were
obtained from the fit to local masses of smeared-smeared correlators.
They were independent on the parameters $\alpha$ and $n$ of the
applied smearing function (24)
\footnote{Although the t-dependent local masses depend on the
underlying wave function, they finally end up with the same plateau!}.
Having established the wave function independence, for the rest of the
calculation we performed the estimation of the spectrum using only the
optimized wave function with $\alpha=3$ and $n=25$.\\
The masses were calculated at different $k<k_c$'s
from a two parameter fit to eq. (22) or (34) on a
time interval $[t_{min},t_{max}]$ determined by the plateaux and by studying
the stability of the fits under changes to the fitting range. The tipical best
fitting ranges
for the different lattices A,B,C (see Table 1) are
$[t_{min},t_{max}]=$[6,10], [4,7] and [10,16],
respectively.\\
All needed fits
were performed by minimizing the correlated $\chi^2$ and the statistical errors
were calculated using the jackknife method \cite{16} (with binning to control
the
autocorrelation).

\subsection{Phase diagram and $a^{-1}(\beta)$ dependence}
The critical hopping
parameter $k_c$ was determined by linearly extrapolating $m_{\gamma_5}^2$ to
the chiral limit as explained in section 3.3. For the different lattices
the phase diagram in the $(k,\beta)$-plane is plotted in Fig. 1.
In this range of the parameter space there are
no evident finite volume and $a$ effect for these data. Data obtained
from simulations with topological trivial and topological non-trivial
configurations are consistent with each other.\\
In a previous
work \cite{13} we have computed the mass $m_{\gamma_5}$ by a strong coupling
expansion to high orders. In Fig. 2 we plot in the $(k,\beta)$-plane
the critical line determined by the
chiral limit $m_{\gamma_5}(k,\beta)=0$ over a large range of $\beta$ values
(from 0.0 to 2.8).  In the same plot we compare the critical line
with the Monte Carlo data.
Considering that the convergence radius of the strong coupling expansion can
be estimated to be $\beta_{conv.}\simeq 1.7$
the agreement between the two calculation is surprising. \\
The lattice spacing $a$ was determined from the vector isotriplet mass
extrapolated to the chiral limit and normalized with the experimental value of
the W-boson mass. In Fig.
3 we have plotted the Monte carlo data for the different lattices.
We clearly see in the plot that the values of
$a^{-1}(\beta)$ for a fixed $\beta$
are affected by finite volume effects, as expected.
The prediction of the strong coupling expansion and the Monte Carlo data are
plotted in Fig. 4. For values of $\beta$ less than the estimated convergence
radius $\beta_{conv.}\simeq 1.7$ we obtain the function $a^{-1}(\beta)$ (solid
line). For values of $\beta$ bigger than $\beta_{conv.}$
the behaviour of this function can be interpolated by
the Monte Carlo data (dashed line).

\subsection{Spectrum}
\subsubsection{The axial vector and tensor bound state masses}
The masses of the axial vector $V^k$ and the
tensor $\tilde B^{kjA}$ bound states are evaluated on the lattices Ai
(i=1,...3), Bi (i=0,..3) and Ci (i=1,..3) (see Table 1) using topological
trivial gauge configurations. After a thermalization by
5000 iterations sweeps we use gauge
configurations separated by 200 sweeps.
For 1 heat bath sweeps we do 6 over-relaxed sweep.\\
For each $\beta$ we have fitted the
masses using eq. (22) for eight different hopping parameters and extrapolated
them to the chiral limit.\\
In Fig. 5 and Fig. 6 we present the result of the lattices B1, B2, B3 and C1,
C2, C3 for
which the finite volume effects are small. The same results are presented also
in
Table 2.
We plot the masses as a function of the lattice spacing $a$
normalized with the W-boson mass at 80 GeV. All data points with
$m_\Gamma\times a>1$ are excluded from the analysis of the finite $a$ effects.
These correspond to masses evaluated on the lattice B0
with the smallest $\beta$.
The continuum limit of the
masses is evaluated by linearly extrapolating to $a=0$ for the different
lattices. The results of the extrapolations are presented in Fig. 5 and Fig. 6.
Finite volume effects at the continuum limit are negligible between the
lattices B1, B2, B3 and C1, C2, C3 in comparison
to the corresponding statistical errors.
\subsubsection{The pseudoscalar isosinglet mass}
The mass of the pseudoscalar isosinglet $S_-^{an}$ is evaluated on the
lattices B0*, B1* and B2*. We used topological non-trivial gauge configurations
with topological charge $Q=1$. After a thermalization by
5000 iterations sweeps we used gauge
configurations separated by 1000 sweeps. The evolution of the
topological charge during the iterations is plotted in Fig. 7. \\
For each $\beta$ we have fitted the mass difference $\Delta m_{\gamma_5}$
with the ratio $R(t)$ of eq. (34)
for five different hopping parameters $k$.
In Fig. 8 we plot the different values of $\Delta m_{\gamma_5}\times a$ as a
function of the
$m_{\gamma_5}^2\times a^2$ (which is proportional to the inverse hopping
parameter).
The same result is presented also in Table 3.
We note that it is
independent of $m_{\gamma_5}^2$, as it should be. After checking this mass
independence
the extrapolation to the chiral limit at $m_{\gamma_5}^2=0$ was done by a
constant
function.\\
In Fig. 9 we plot the chiral limit of $\Delta m_{\gamma_5}$ as a function
of the lattice spacing $a$. At the chiral limit the mass difference coincides
with
the mass of the pseudoscalar isosinglet $m_{\gamma_5}^{an}$.
We clearly see that decreasing the lattice spacing $a$ the pseudoscalar
isosinglet mass increase and it is bigger than the W-boson mass.
\section{Conclusion}
We have discussed a composite model for the weak bosons of the SM.
The model is based on a Yang-Mills theory without Goldstone bosons.
In this theory the W-boson of the SM is represented by the vector isotriplet
bound state.
As a main result
our lattice Monte Carlo simulation has shown that the vector
isotriplet bound state is the lightest bound state in our model as it should be
for any viable candidate of electroweak composite model.\\
We have developed an efficient method to compute the mass of the pseudoscalar
isosinglet bound states. This computation requires the evaluation of
disconnected fermion loops and the generation of topological non-trivial gauge
configurations.\\
We have also predicted the mass of the first bound states heavier than the
vector isotriplet. These bound states are a vector isotriplet
and a vector isosinglet and a pseudoscalar with
masses in the range of a few hundred GeV (see Fig. 5, Fig. 6 and Fig. 9).
These predictions open new experimental perspective at LEPII and LHC.\\[1cm]
{\Large {\bf Acknowledgements}}\\[0.5cm]
We would like to thank  C.Alexandrou, F.Jegerlehner and
H.Schlereth for their
help,  Ph. de Forcrand from whom we have obtained the program to generate the
gauge configurations and F.Jegerlehner for reading and correcting the
manuscript.

\vspace{0.5cm}
{\Large {\bf Figure Caption}}
\begin{enumerate}
\item Lattice Monte Carlo data for the phase diagram in the $(\beta,k)$-plane.
\item Lattice Monte Carlo data and strong coupling expansion prediction for the
phase
diagram in the $(\beta,k)$-plane. The dotted line represents the limit of 1/8
for $\beta\rightarrow\infty$.
\item Lattice Monte Carlo data for
the inverse lattice spacing $a^{-1}$ as a function of $\beta$.
\item Lattice Monte Carlo data and strong coupling expansion calculation
for the inverse lattice space $a^{-1}$ as a function of $\beta$. The dotted
line
represents the estimate convergence radius of the strong coupling expansion.
The
dashed line is a qualitative estimation of $a^{-1}(\beta)$ based on an
interpolation of the Monte Carlo data of the biggest lattice.
\item Lattice Monte Carlo prediction of the triplet $\tilde B_{jk}^A$
bound state mass as a function of the lattice spacing. Dashed
and dotted lines are the extrapolation to the continuum limit.
\item Lattice Monte Carlo prediction of the axial vector $V^k$ bound
state mass as a function of the lattice spacing. Dashed
and dotted lines are the extrapolation to the continuum limit.
\item Evolution of the topological charge during the iterations. The
topological charge is measured at each 50 iterations.
\item The mass difference $\Delta m_{\gamma_5}^{an}$ as a function of the
$m_{\gamma_5}^2$ mass in unit of the lattice spacing $a$. The dashed lines
represent the extrapolations to the chiral limit.
\item The pseudoscalar isosinglet mass as a function of the lattice spacing
$a$.
\end{enumerate}
\newpage
\begin{table}
\begin{tabular}{|c|c|lr|c|c|c|}\hline
lattice & $\beta$ & $N_s$,&$N_t$ & no. config's. &$a^{-1}/GeV$&$k_c$\\\hline
A1      &  2.3    & 6,&24& 44 & 120(10)&0.164(6)\\
A2      &  2.4    & 6,&24& 40 & 145(12)&0.156(6)\\
A3      &  2.5    & 6,&24& 40 & 156(24)&0.153(4)\\\hline
B0*     &  2.2    & 8,&16& 40 & 129(7)&0.173(1)\\
B0      &  2.2    & 8,&16& 40 & 119(9)&0.173(1)\\
B1*     &  2.3    & 8,&16& 40 & 156(4)&0.170(2)\\
B1      &  2.3    & 8,&16& 40 & 141(8)&0.166(2)\\
B2*     &  2.4    & 8,&16& 40 & 181(17)&0.154(2)\\
B2      &  2.4    & 8,&16& 42 & 174(16)&0.154(1)\\
B3      &  2.5    & 8,&16& 40 & 198(17)&0.153(2)\\
B4      &  2.7    & 8,&16& 40 & *******&0.152(8)\\\hline
C1      &  2.3    & 12,&36& 40 & 158(6) &0.1672(4)\\
C2      &  2.4    & 12,&36& 40 & 177(4) &0.1558(3)\\
C3      &  2.5    & 12,&36& 40 & 225(12)&0.1515(4)\\\hline
\end{tabular}
\caption{Parameters of the lattices used for this work. The inverse lattice
spacing is obtained by fixing the vector isotriplet mass to 80 GeV. The
critical hopping parameter is obtained by extrapolating $m_{\gamma_5}$ to the
chiral limit. Asterisks (*****) indicate that the fit was not accepted due to
a large $\chi^2$ or the missing of a plateau in the local masses in the region
of the fit. The lattices B0*, B1* and B2* indicate simulations performed with
topological non-trivial configurations.}
\end{table}

\newpage
\begin{table}
\begin{tabular}{|c|c|c|c|}\hline
lattice & $a^{-1}$/GeV & $m_{\sigma^{jk}T^I}$ GeV& $m_{\gamma_5\gamma^k}$ GeV
\\\hline\hline
A1      & 120(10)      & 73(47) & 83(48)\\
A2      & 145(12)      & 76(21) & 104(27)\\
A3      & 156(24)      & 111(40)& 98(39)\\\hline
B0      & 119(9)       & 202(36)& 152(37)\\
B1      & 141(8)       & 108(27)& 125(24)\\
B2      & 174(16)      & 135(25) & 130(31)\\
B3      & 198(17)      & 122(26) & 128(29)\\\hline
C1      & 158(6)       & 97(14)  & 128(27)\\
C2      & 177(4)       & 96(10)  & 163(34)\\
C3      & 225(12)      & 113(20) & 133(20)\\\hline
\end{tabular}
\caption{Lattice Monte Carlo predictions of the triplet $\tilde B_{jk}^I$ and
singlet $V^k$ bound states as a function of the inverse
lattice spacing $a^{-1}$. }
\end{table}

\begin{table}
\begin{tabular}{|c|c|c|c|}\hline
lattice & $a^{-1}$/GeV &  $m^2_{\gamma_5}\times a^2$&
$\Delta m^{an}_{\gamma_5}\times a$ \\\hline\hline
B0*    & 129(7) &0.270(27) & 0.71(15)\\
       &        &0.359(20)& 0.71(14)\\
       &        &0.442(18)& 0.70(14)\\
       &        &0.562(16)& 0.69(12)\\
       &        &0.719(17)& 0.68(11)\\\hline
B1*    &156(4)  &0.489(6) & 0.98(15)\\
       &        &0.543(7) & 1.00(16)\\
       &        &0.596(7) & 1.02(17)\\
       &        &0.632(8) & 1.03(17)\\
       &        &0.702(9) & 1.05(18)\\\hline
B2*    & 181(17)&0.535(14)& 0.97(9)\\
       &        &0.610(14)& 1.00(10)\\
       &        &0.712(13)& 1.06(12)\\
       &        &0.776(12)& 1.10(13)\\
       &        &0.830(11)& 1.14(14)\\\hline
\end{tabular}
\caption{Lattice Monte Carlo predictions of the mass difference
$\Delta m^{an}_{\gamma_5}$ in unit of the lattice spacing.}
\end{table}

\end{document}